\documentclass{elsart}

\usepackage{natbib}

\usepackage{epsfig}

\begin{document}

\begin{frontmatter}



\title{Astrometric effects of solar-like magnetic activity in late-type stars
 and their relevance for the detection of extrasolar planets}


\author[INAF-CT]{A.~F.~Lanza\corauthref{cor}},
\ead{nuccio.lanza@oact.inaf.it}
\author[INAF-CT,UNI-CT]{C.~De Martino},
\author[INAF-CT,UNI-CT]{M.~Rodon\`o\thanksref{tha}}

\address[INAF-CT]{INAF-Osservatorio Astrofisico di Catania,}
\address[UNI-CT]{Dipartimento di Fisica e Astronomia, Universit\`a degli Studi di Catania, Via S.~Sofia, 78, 95123 Catania, Italy}

\corauth[cor]{Corresponding author}

\thanks[tha]{Deceased}

\begin{abstract}
Using a simple model based on the characteristics of sunspots and faculae in solar active regions,
the effects of surface brightness inhomogeneities on the position of the photocentre 
of the disk of a solar-like, magnetically active star, are studied. A general law is introduced, giving the maximum
amplitude of the photocentre excursion produced by a distribution of active regions with a given
surface filling factor.  The consequences for the detection
of extrasolar planets by means of the astrometric method are investigated with some 
applications to forthcoming space missions, such as GAIA and SIM, as well as to 
ground-based interferometric measurements. Spurious detections of extrasolar planets can indeed be caused
by activity-induced photocentre oscillations, requiring a simultaneous monitoring of the
optical flux and a determination of the rotation period and of the level of activity of the 
target stars for an appropriate discrimination. 

\end{abstract}


\end{frontmatter}

\section{Introduction}
\label{intro}

The distribution of brightness over the solar photosphere is not uniform, due to the
presence of dark sunspots and bright faculae in active regions and to the  
granular convection and bright network elements, associated with slender flux tubes located in the
intergranular lanes. The baricentre of the brightness distribution over the solar disk, i.e.,
the photocentre of the Sun, is affected by the brightness inhomogeneities having the 
largest scales, that is by sunspots and faculae,  because the averaged effect of the 
granulation and  the network elements is negligible, thanks to
their almost uniform distribution on spatial scales comparable with the solar radius. 
The rotation of the Sun and the evolution of the active regions make the perturbation of the
position of the photocentre a function of the time. 

Sunspots and faculae are manifestations of the magnetic field, amplified and modulated 
by the solar dynamo, and are present also on the photospheres of late-type stars
hosting hydromagnetic dynamos in their convective envelopes. Therefore, the brightness
inhomogeneities observed in the Sun are also present on late-type stars during their
main-sequence lifetime and are most prominent in young, rapidly rotating stars, given their
more powerful dynamo action
\citep[e.g. ][]{LanzaRodono99}.  

The perturbation of the position of the photocentre of late-type stars due to their
magnetic activity {should be} considered in some detail, 
in view of the forthcoming space missions GAIA\footnote{See http://www.rssd.esa.int/Gaia} 
and SIM\footnote{See http://sim.jpl.nasa.gov/}. 
{ They will obtain 
astrometric measurements of the position of late-type stars with
$V \leq 10-12$  with an accuracy between $1$ and $10$ $\mu$as (micro arc-seconds); see, e.g., 
\citet{Shaoetal07}}. Moreover, 
ground-based interferometric instruments, such as PRIMA at  
VLTI\footnote{See http://www.eso.org/projects/vlti/instru/prima/index\_prima.html},
will also reach an accuracy of $\sim  10$ $\mu$as in the determination of the photocentre
position. { The feasibility of such a ground-based high-precision differential astrometry 
has been recently demonstrated by the PHASES (Palomar High-precision Astrometric Search 
for Planets) program \citep[see][]{Muterspaughetal06}. }

In the present paper, a simple model for computing the astrometric effects of
cool spots and bright solar-like faculae in 
a late-type star is introduced, and applied to derive the maximum expected
amplitude of variation of the photocentre position as a function of the active region filling factor.
Since the  perturbation is cyclically modulated by the stellar rotation, 
it can be a source of spurious detections of close-by planetary 
companions by means of the astrometric technique. We present some cases leading 
to possible confusion and propose methods to discriminate the effects 
of a planet from those due to stellar activity. 

\section{The model}
\label{model}

Let us consider a Cartesian reference frame with the $\hat{x}$-$\hat{y}$ plane 
in the plane of the sky, that is the
origin O at the unperturbed photocentre of the star, the $\hat{x}$-axis 
along the declination direction pointing to the North pole, the $\hat{y}$-axis along the 
direction of the increasing right ascension and the $\hat{z}$-axis pointing toward the observer.
The position $(x_{\rm p}, y_{\rm p})$ of the photocentre 
at a given time $t$ depends on the  distribution of the brightness on
the surface of the apparent disk of the star at the isophotal wavelength $\lambda$ of the observations, and is given by:
\begin{equation}
\begin{array}{l}
x_{\rm p} = \frac{\displaystyle \int_{D} I(\lambda, x, y) x dS}{\displaystyle \int_{D} I (\lambda, x, y) dS}, \\
 \mbox{}  \\
y_{\rm p} = \frac{\displaystyle \int_{D} I(\lambda, x, y) y dS}{\displaystyle \int_{D} I (\lambda, x, y) dS}, \\
\end{array}
\label{pc_def}
\end{equation}
where $I$ is the specific intensity at wavelength $\lambda$ at the given position $(x, y)$ on the stellar disk, 
and the integration is
extended over the apparent stellar disk $D$. Equations (\ref{pc_def})
define the photocentre as the baricentre of the flux distribution emerging from the
stellar disk. For an unperturbed star, such a distribution is symmetric around the
disk centre and the photocentre coincides with the projection of the baricentre of 
the star on the plane of the sky. 

In order to describe the distribution of the surface brightness when there are
photospheric inhomogeneities, it is more convenient to adopt a spherical 
reference frame rotating with the star. 
Specifically, a spherical reference
frame is considered with the origin at the baricentre O$_{\rm B}$ of the star and the $\hat{z_{0}}$-axis
directed along its rotation axis, rotating with the stellar angular velocity  
$\Omega$ with respect to an inertial reference frame. 
We can express the position of a generic point on the stellar surface
by means of its Cartesian co-ordinates in that frame:
\begin{equation}
\left\{ 
\begin{array}{l}
x_{0} = {\cal R} \sin \theta \cos l, \\
y_{0} = {\cal R} \sin \theta \sin l, \\
z_{0} = {\cal R} \cos \theta, \\ 
\end{array}
\right.
\label{solidal_coo}
\end{equation}
where ${\cal R}$ is the radius of the star, assumed to be spherically symmetric, and $\theta$ and $l$ are the colatitude and the 
longitude of the point on the stellar surface, respectively. The transformation
from the latter  reference frame to the reference frame of the
stellar disk can be achieved by means of three successive Eulerian rotations.
In order to define them, let us consider that the stellar equatorial plane 
intersects the plane of the sky along a line of nodes that makes an angle $\Phi$
with respect to the direction of the $\hat{x}$-axis on the plane of the sky. 
The first rotation by an angle $\omega = - \Omega (t -t_{0})$ around the $\hat{z}_{0}$-axis
makes the fundamental stellar meridian coincide with its position at the time $t_{0}$ when 
it crosses the line of nodes. The second rotation by the inclination $i$ of the
stellar rotation axis, around the line of nodes, makes the $\hat{z}_{0}$-axis coincide 
with the $\hat{z}$ axis. Finally, a rotation by an angle $\Phi$, around the $\hat{z}$ 
axis, reports the reference frame into that of the plane of the sky. 

In matrix form,  we can write the co-ordinate transformation as follows:
\begin{equation}
\left[ 
\begin{array}{l}
x \\
y \\
z \\
\end{array}
\right] = \tilde{A} 
\left[ 
\begin{array}{l}
x_{0} \\
y_{0} \\
z_{0} \\
\end{array}
\right], 
\label{transf}
\end{equation}
where $\tilde{A}$ is the transformation matrix that is given by: 
\begin{equation}
\tilde{A} = 
\left(
\begin{array}{ccc}
\cos \Phi & \sin \Phi & 0 \\
-\sin \Phi & \cos \Phi & 0 \\
0 & 0 & 1 \\
\end{array}
\right)
\left(
\begin{array}{ccc}
1 & 0 & 0 \\
0 & \cos i & \sin i \\
0 & -\sin i & \cos i\\
\end{array}
\right)
\left(
\begin{array}{ccc}
\cos \omega & \sin \omega & 0 \\
-\sin \omega & \cos \omega & 0 \\
0 & 0 & 1 \\
\end{array}
\right), 
\end{equation}
where the individual factor matrixes express the three rotations as specified above, 
respectively. Note that the angle $\omega$ is a linear function of the time, whereas $i$ and $\Phi$ are constant. 

We  subdivide the photosphere of the star into $1^{\circ} \times 1^{\circ}$
elements, each of which is identified by the longitude $l$ and colatitude $\theta$ of its central point.
The status of the photosphere in each element can be either unperturbed, spotted, or of facular type.
The specific intensity of the unperturbed photosphere is assumed to be given by a linear limb-darkening law:
\begin{equation}
I(\lambda, \mu) = I_{0}(\lambda) ( 1 -u_{\lambda} + u_{\lambda} \mu), 
\end{equation}
where $I_{0}(\lambda)$ is the specific intensity at the centre of the disk,  
$u_{\lambda}$  the linear limb-darkening coefficient at the isophotal wavelength $\lambda$,
and $\mu \equiv \cos \psi$, where $\psi$ is the angle between the normal to the given surface element
and the line of sight. It is given by: 
\begin{equation}
\mu = \sin \theta \sin i \cos \left[ l + \Omega (t -t_{0}) \right] + \cos \theta \cos i. 
\end{equation}
The specific intensity of the spotted photosphere is given by: $I_{\rm s} (\lambda, \mu) = c_{\rm s}(\lambda) I(\lambda, \mu)$, where 
the spot contrast $c_{\rm s} < 1$ is a function of the wavelength $\lambda$ and is assumed to be independent
of $\mu$ \citep[cf. ][]{Lanzaetal04}. The specific intensity of the facular photosphere is assumed to be equal
to that of the unperturbed photosphere at the disk centre, while its contrast increases linearly toward the limb,
i.e., $I_{\rm f}(\lambda, \mu) = I(\lambda, \mu) [1 + c_{\rm f}(\lambda) (1 - \mu)]$, with 
the constrast coefficient $c_{\rm f}$ being a function of the
wavelength \citep[see ][ for details]{Lanzaetal04}. The value of the flux  coming from the $i$-th surface
element at the wavelength $\lambda$ and the time 
$t_{k}$ is given by:
\begin{equation}
\Delta F_{i} (\lambda, t_{k}) = \left\{ 
   \begin{array}{ll}
        I[\lambda, \mu(t_{k})] \Delta A_{i} \mu(t_{k}) g[\mu(t_{k})] & \mbox{ for the unperturbed photosphere,} \\
        I_{\rm s}[\lambda, \mu(t_{k})] \Delta A_{i} \mu(t_{k}) g[\mu(t_{k})] & \mbox{ for the spotted photosphere,} \\
         I_{\rm f}[\lambda, \mu(t_{k})] \Delta A_{i} \mu(t_{k})  g[\mu(t_{k})] & \mbox{ for the facular photosphere,}\\
   \end{array}
   \right. 
\end{equation}
where $g$ is the visibility function of the surface element, i.e., $g(\mu) = 1$ for $\mu \geq 0$, $g(\mu) = 0$ for $\mu <0$.
The total flux from the stellar disk
at time $t_{k}$ is: $F(\lambda, t_{k}) = \sum_{i} \Delta F_{i} (\lambda, t_{k})$.  
The co-ordinates of the photocentre on the plane of the sky at time $t_{k}$ are given by:
\begin{equation}
\begin{array}{ccc}
x_{\rm p}  & = & \frac{\displaystyle \sum_{i} x_{i} (t_{k}) \Delta F_{i} (\lambda, t_{k})}{\displaystyle \sum_{i} 
\Delta F_{i} (\lambda, t_{k})}, \\ 
y_{\rm p}  & = & \frac{\displaystyle \sum_{i} y_{i} (t_{k}) \Delta F_{i} (\lambda, t_{k})}{\displaystyle \sum_{i} 
\Delta F_{i} (\lambda, t_{k})}, \\ 
\end{array}
\label{ph_res}
\end{equation}
where the co-ordinates $x_{i}$ and $y_{i}$ depend on the spherical co-ordinates $l_{i}, \theta_{i}$ of the $i$-th surface element 
and the geometrical parameters of the star through Eqs. (\ref{solidal_coo}) and (\ref{transf}). From Eq.~(\ref{ph_res}), we see that
$x_{\rm p}$ and $y_{\rm p}$ depend on the isophotal wavelength of the observations. It can be assumed to be fixed for a given
astrometric instrument, hence we shall not consider explicitly that dependence.   

Other effects affecting the position of the photocentre, e.g., the oscillations
due to a planet or a planetary system \citep[e.g. ][]{Sozzettietal03}, or the proper motion of the star, can be
linearly added up to the effects of the surface brightness inhomogeneities computed with the scheme introduced above.

\section{Results}
\label{results}

\begin{figure}
\epsfig{file=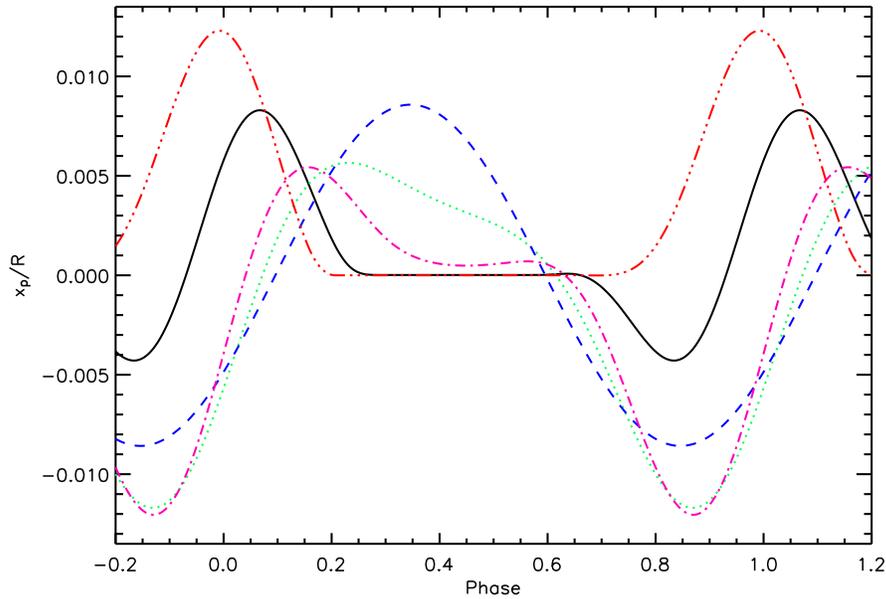,width=12cm}
\caption{The relative abscissa $x_{\rm p}$ of the photocentre of our model star on the plane of the sky versus the rotational phase.
The unit of measure of the abscissa is the apparent radius $R$ of the stellar disk.
Different linestyles and colors  refer to different values of the inclination $i$ of the rotation axis: dashed blue -- $i=0^{\circ}$ (i.e., pole-on view); dotted green -- $i=15^{\circ}$; dot-dashed pink -- $i=30^{\circ}$;  solid black -- $i=60^{\circ}$; and
three-dot-dashed red -- $i=90^{\circ}$ (i.e., equator-on view).  }
\label{phot_ex}
\end{figure}

\begin{figure}
\epsfig{file=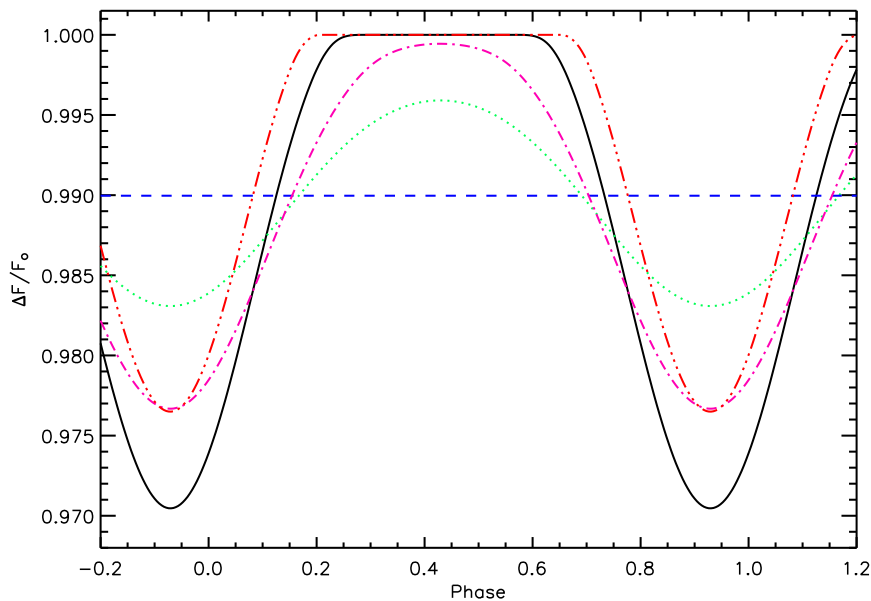,width=12cm}
\caption{The relative variation of the stellar flux as a function of the rotational 
phase for the cases simulated in Fig.~\ref{phot_ex}. The unit of flux $F_{0}$ corresponds to the unperturbed flux of the star. 
Different linestyles and color indicate different values of the inclination, as in Fig.~\ref{phot_ex}. }
\label{flux_var}
\end{figure}

We have applied the model described in Sect.~\ref{model} to compute the apparent motion of the photocentre of a star
with a single active region on its surface, containing only cool spots and having a longitudinal extension of $20^{\circ}$ 
between latitudes $20^{\circ}$ and $40^{\circ}$ in the Northern hemisphere. 
{ The active region has an area of 0.83\% of the whole stellar surface, i.e., about $\sim 8-10$ times 
larger than the area of the largest  sunspot groups at the maximum of the eleven-year cycle. 
However, we believe this is more appropriate for late-type stars that are slightly more active than the Sun 
(see Sect.~\ref{application}. 
For the sake of simplicity, our model spotted region is assumed to be uniform in contrast and  squared in shape (see 
Fig.~\ref{figAR}). However, as we shall see below,  the maximum photocentre excursion depends primarily
on the total area of the active region with its shape playing a secondary role. 
}
\begin{figure}
\epsfig{file=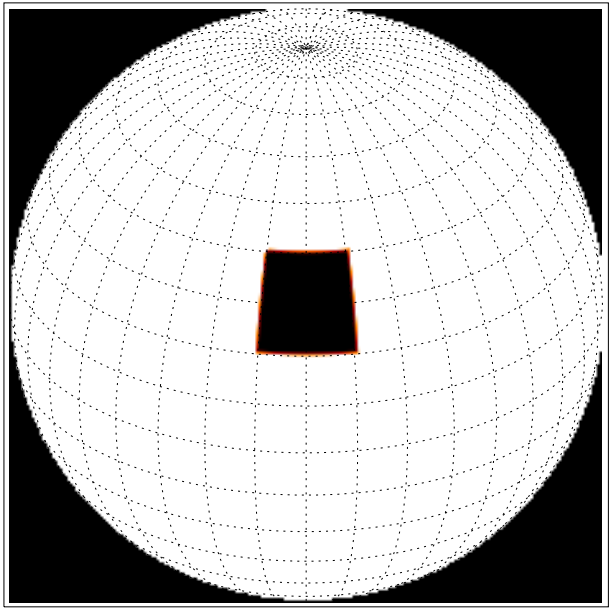,width=12cm}
\caption{ The disk of our model star with the spotted region of $20^{\circ}\times 20^{\circ}$ as seen by 
the observer at the rotation phase when the active region transits across the central disk meridian.
 The inclination of the stellar rotation axis with respect to the line of sight is $60^{\circ}$, thus the
centre of the active region passes through the centre of the star disk.   }
\label{figAR}
\end{figure}
The limb-darkening coefficient of the
unperturbed photosphere is assumed to be $u_{\lambda} = 0.65$ at the given isophotal wavelength and the spot contrast 
$c_{\rm s} = 0.33$, which is appropriate to describe the bolometric flux perturbation in the case 
of sunspots,  
{ corresponding to an effective temperature of the spotted photosphere of $\sim 4400$ K, i.e., $\sim 1400$ K 
below that of the unperturbed photosphere, which accounts, on  average, for the combined effects 
of umbrae and penumbrae \citep[e.g., ][]{Chapmanetal94}}.
 In Fig.~\ref{phot_ex}, 
we plot the position of the photocentre versus the rotational phase for different values of the 
inclination $i$ of the stellar rotation
axis along the line of sight. The photocentre excursion is measured in unit of the apparent radius of the stellar disk
$R$ along the direction of the $\hat{x}$-axis, adopting a position angle $\Phi = 30^{\circ}$ in our calculations.
The motion along the $\hat{y}$-axis is similar, but shifted in phase by an amount 
that depends on the co-ordinates of the active region and the stellar parameters $i$ and $\Phi$. 
 The active region is
assumed to be stable, so that stellar rotation is the only effect responsible for the motion of the 
photocentre. 
The variations of the relative flux corresponding to the simulations shown in Fig.~\ref{phot_ex} are plotted in Fig.~\ref{flux_var}.
When the inclination is lower than $20^{\circ}$, the active region is always in view and produces a continuous 
modulation of the position of the photocentre and of the flux that depends on the variation of its position 
on the stellar disk and  its projected area. For a pole-on view, the projected area is constant and the
variation of the photocentre abscissa is purely sinusoidal, whereas the flux stays constant. When the inclination increases, the 
modulation of the photocentre abscissa is no longer sinusoidal. When $i \geq 50^{\circ}$, there are 
 intervals of phase during which the photocentre
is unperturbed, i.e., when the active region transits on the invisible hemisphere of the star.
 Considering the case with $i=60^{\circ}$, 
we note that, when the active region first appears 
on the rising part of the disk, it produces  a negative
shift of the photocentre abscissa because the baricentre of the flux distribution falls in the opposite half of the disk. 
Then the shift reverses its sign, when the active region is carried by stellar rotation on the other half of the disk, and eventually
it becomes zero when the active region comes out of view. For $i\not=90^{\circ}$, the duration of the active region 
transit across the disk depends on its average latitude and the inclination. Moreover, the sign and the 
value of the photocentre excursion
depend also on the position angle of the projection of the stellar rotation axis on the plane of the sky, as measured by the
angle $\Phi$. The combination of these different factors explains the different shapes and amplitudes of the curves 
plotted in Fig.~\ref{phot_ex}.  

\begin{figure}
\epsfig{file=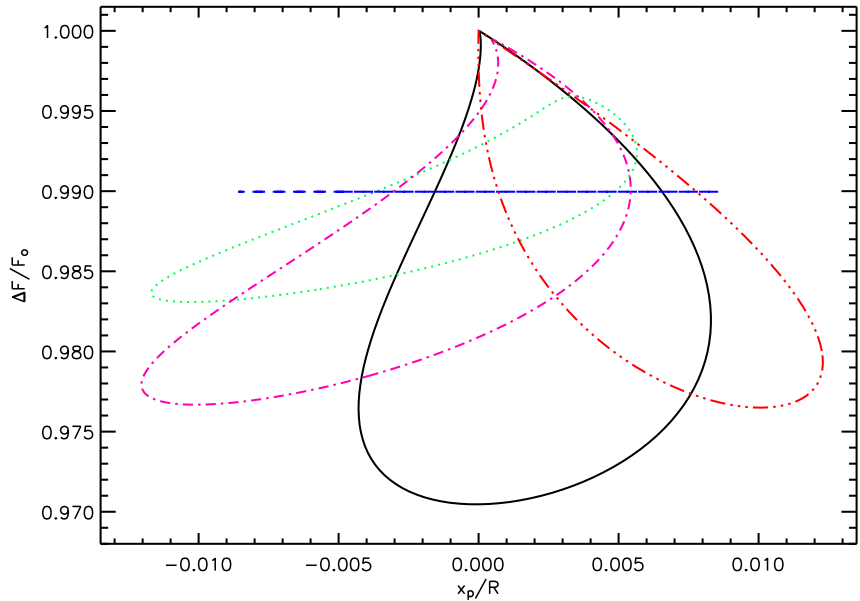,width=12cm}
\caption{The relative flux variation versus the photocentre abscissa for the cases simulated in Fig.~\ref{phot_ex}. 
Different linestyles and colors indicate different values of the inclination, as in Fig.~\ref{phot_ex}. Note the correlation between 
the photocentre motion and the flux variation, and the degenerate case corresponding to $i=0^{\circ}$.}
\label{correl}
\end{figure}

The flux variation produced by the active region increases 
with increasing inclination and reaches its maximum value 
when the active region transits through the centre of the stellar disk, i.e., for $i=60^{\circ}$, because in that case its
projected area reaches its maximum value. Note that the excursion of the photocentre and the flux variations are
correlated for $i > 0^{\circ}$, as shown in Fig.~\ref{correl}. 

\begin{figure}
\epsfig{file=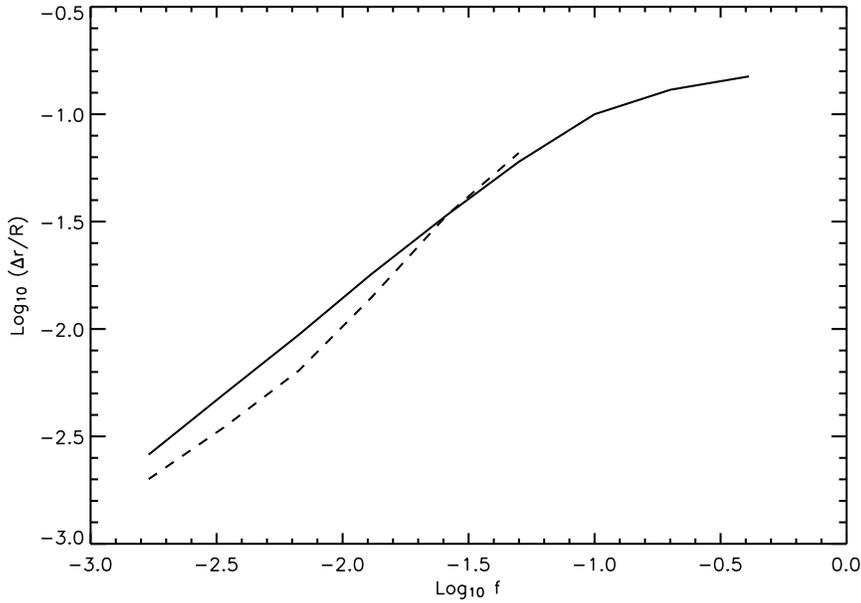,width=12cm}
\caption{The relative excursion $\Delta r/R$ of the photocentre  versus the 
spot filling factor $f$ in the case of a single active region for an inclination of the stellar rotation axis of $i=60^{\circ}$.
The solid line refers to the case of an active region consisting only of dark spots, whereas the dashed line refers to an
active region with dark spots surrounded by solar-like faculae (see the text).  }
\label{excur_l}
\end{figure}

When only a single active region is present on the stellar disk, the excursion of  the photocentre reaches its maximum
value for a given value of the total active region area. As a matter of fact, 
if the same total area is subdivided into two or more active regions,
the excursion generally decreases because the simultaneous presence of more than one brightness inhomogeneity leads to
a reduction of their combined effects.  Therefore, we have investigated the variation of the maximum photocentre excursion 
$\Delta r \equiv \max\left\{\sqrt{(\Delta x_{\rm p})^{2} + (\Delta y_{\rm p})^{2}} \right\}$ as a function of the filling factor $f$ 
for a single active region by means of our model (see Fig.~\ref{excur_l}). 
The stellar inclination is assumed
fixed at $i=60^{\circ}$, that is the most probable value for  a random orientation of the stellar rotation axis. 
The average latitude of the active region is $30^{\circ}$ to maximize its effect on the position of the photocentre. 
In those simulations, we consider also the effect of a facular component that surrounds the dark spot. The area of 
the facular component is assumed to be ten times that of the cool spots, as suggested 
by solar observations and modelling of the solar irradiance variations 
\citep[see, e.g., ][ and references therein]{Lanzaetal03,Lanzaetal07}. The spot constrast is assumed to be 
$c_{s} = 0.33$, while the facular contrast is $c_{\rm f} = 0.15$ \citep[see, e.g., ][]{Lanzaetal07}.
Note that the effect of the faculae is small, in spite of their relative  area being ten times that of cool spots.
This is due to the fact that their effect is significant only close to the limb, 
because their contrast is negligible close to the centre of the disk. 
Moreover, the projection effect disfavours the facular contribution, because their projected area becomes
small close to the limb.  The situation changes if we consider hot spots instead 
of solar-like faculae, that is bright regions the constrast
of which is significant also close to the disk centre, as is sometimes observed in the case of very active stars \citep[cf., e.g.,
the case of the active component of { the RS CVn binary system} HR 1099 reported by ][]{Vogtetal99}. However, since we are here mainly interested in
solar-like stars, we shall not consider further this possibility. 
\begin{figure}
\epsfig{file=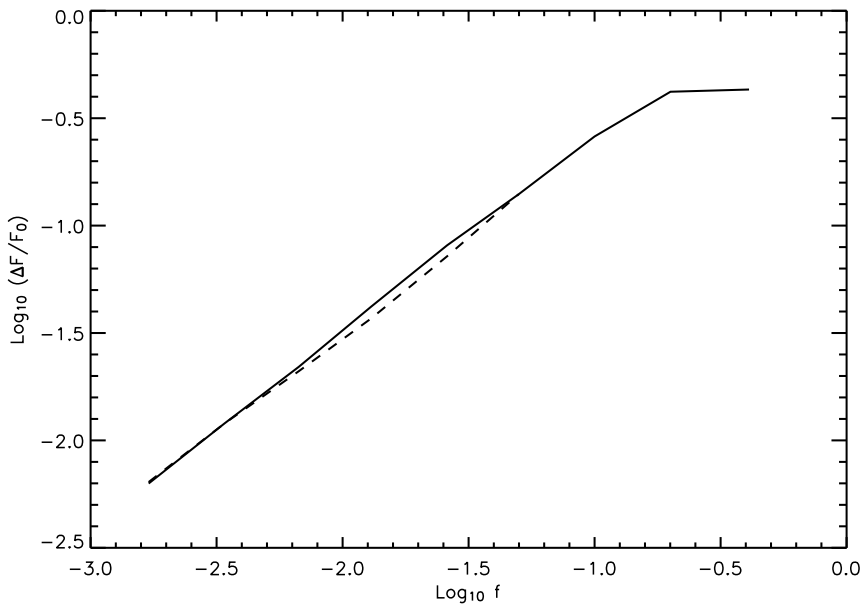,width=12cm}
\caption{The relative flux variation $\Delta F/F_{0}$ versus the 
spot filling factor $f$ in the case of a single active region for an inclination of the stellar rotation axis of $i=60^{\circ}$.
The solid line refers to the case of an active region consisting only of dark spots, whereas the dashed line refers to an
active region with dark spots surrounded by solar-like faculae (see the text).  }
\label{flux_l}
\end{figure}

The variation of the amplitude of the rotational modulation of the stellar flux versus the filling factor is plotted in 
Fig.~\ref{flux_l}. The correlations shown in Figs.~\ref{excur_l} and \ref{flux_l} can be well approximated by power laws
up to a value of the filling factor $f \simeq 0.1$, beyond which they deviate from such a simple dependence because the
amplitude of the variation saturates when a large fraction of the stellar disk is covered with active regions. 
Applying a linear best fit to the  plots in Figs.~\ref{excur_l} and \ref{flux_l}, we find the following 
regression expressions, valid for $ f\leq 0.1$, with a correlation coefficient $ \rho > 0.99$:
\begin{eqnarray}
\log_{10} \frac{\Delta r}{R}  & = & (-0.055 \pm 0.031) + (0.907 \pm 0.016) \log_{10} f \mbox{\small (dark spots only),} \nonumber \\
 \log_{10} \frac{\Delta r}{R} & = & (0.176 \pm 0.089) + (1.059 \pm 0.042) \log_{10} f \mbox{\small (spots \& faculae),} 
\label{power_law}
\end{eqnarray}
and 
\begin{eqnarray}
\log_{10} \frac{\Delta F}{F_{0}}  & = & (0.343 \pm 0.014 ) + (0.917 \pm 0.007) \log_{10} f 
\mbox{\small (dark spots only),} \nonumber \\
 \log_{10} \frac{\Delta F}{F_{0}} & = & (0.296 \pm 0.034) + (0.902 \pm 0.017) \log_{10} f \mbox{\small (spots \& faculae),} 
\label{flux_law}
\end{eqnarray}
where the reported uncertainties correspond to one standard deviation. 
The first of Eqs.~(\ref{power_law}) confirms the results previously obtained by \citet{Hatzes02} in the case 
of an active region consisting of dark spots only.
{ \citet{Muterspaughetal06} also estimated upper limits for the astrometric effect of 
dark spots assuming a simplified model
with a fixed projected spot geometry and neglecting  limb-darkening effects. Their model confirms the
almost linear correlation between the variation of the stellar flux and the photocentre excursion that can be 
deduced combining together Eqs. (\ref{power_law}) and (\ref{flux_law}). However,  they predict
an astrometric effect about 60\% larger than ours for a given photometric amplitude, essentially because of 
the neglection of the foreshortening and limb-darkening effects.  }

Note that the simultaneous presence of spots and faculae decreases the photocentre excursion in our model and makes the 
regression more steep, especially for $f \leq 0.03$. This is due to the opposite effects of bright and dark regions
spatially associated on the stellar disk. For a filling factor larger than 0.03, the effect of the faculae becomes negligible
because of their low contrast close to the centre of the disk that leads to a saturation of their contribution. 
 
Equations (\ref{power_law}) and (\ref{flux_law}) can be used to estimate the maximum relative photocentre excursion 
and  flux variation, respectively, expected for a given spot coverage in a solar-like star. 
This is useful to evaluate the possibility that stellar activity may account for an observed 
shift of the photocentre in a late-type star.

\section{Application to nearby stars and consequences for planetary detections}
\label{application}

The effects of stellar magnetic activity on the position of the photocentre 
can be principally observed in nearby solar-like stars, because the maximum excursion cannot exceed the apparent stellar radius.
Considering the one hundred nearest stars listed in the catalogue of \citet{Cox00}, those with spectral type
between F0V and M5V and apparent magnitude $V \leq 11$ have been selected for a quantitative study. They are within 
$\sim 7.6$ pc from the Sun. Their radii 
have been estimated from their $B-V$ color indexes according to \citet{Gray92}, whereas their distances are taken from the 
Hipparcos catalogue. Seven stars turn out to have an apparent radius $R$ 
larger than 1000 $\mu$as and are listed in Table~\ref{table1} where the
columns from the left to the right give the number in the Hipparcos catalogue,
the name, the spectral type, the apparent visual magnitude, the radius, the distance (with its standard deviation), 
and the apparent radius of each star, respectively. 
Moreover, 13 stars turn out to have an apparent radius between $700$ and $1000$ $\mu$as,
while 46 have an apparent radius $ 350 \leq R \leq 700$ $\mu$as. 
\begin{table}
\caption{The seven stars with an apparent radius $R$ larger than 1000 $\mu$as and spectral type later than F5, listed in 
order of decreasing $R$.}
\begin{tabular}{rccrccc}
\hline
   & & & & & & \\
HIP & Name & Sp & $V$   & $\cal{R}$ & $D$ & $R$ \\
    &      &    & (mag) & (${\cal R}_{\odot}$) & (pc) & ($\mu$as) \\
    & & & & & & \\ 
\hline
   & & & & & & \\ 
71683 & $\alpha$ Cen A & G2V & $-0.01$ & 0.91 & $1.347 \pm 0.003$ & 3146 \\
71681 & $\alpha$ Cen B & K1V & 1.35 & 0.76 & $1.347 \pm 0.003$ & 2629 \\
37279 & $\alpha$ CMi & F5IV-V & 0.40 & 1.32 & $3.497 \pm 0.011$ & 1761 \\
87937 & Barnard's star & M5V & 9,54 & 0.57 & $1.821 \pm 0.005$ & 1468 \\
8102  & $\tau$ Cet & G8V & 3.49 & 0.90 & $3.647 \pm 0.011$ & 1153 \\
16537 & $\epsilon$ Eri & K2V & 3.72 & 0.76 & $3.218 \pm 0.009$ & 1101 \\
54035 & Gl 411 & M2Ve & 7.49 & 0.57 & $2.548 \pm 0.006$ & 1041 \\
    & & & & & & \\
\hline
\label{table1}
\end{tabular}
\end{table}
The results obtained in Sect.~\ref{results} show that for a filling factor of 0.2\%, corresponding to the Sun
at the maximum of the eleven-year cycle, excursions as large as $\sim 8$ $\mu$as can be expected for 
$\alpha$ Cen A  in the case of an active region consisting only of 
cool spots. In the case of an active region containing also solar-like faculae, those values are reduced by a
factor of $\sim 1.4$ because the faculae tend to decrease the effect of the spots at such a low value of the filling
factor. A solar value of the filling factor is likely to be appropriate for $\alpha$ Cen A because the star 
appears to have a level of activity comparable or slightly lower than the Sun, with a rotation period of 
about 29 days \citep[see, e.g., ][]{Paganoetal04}. 
Note that the above estimates refer to the case of a single active region, thus the excursion can be reduced if 
a few active regions are simultaneously present on the stellar disk. In any case, those amplitudes will be 
easily detectable by astrometric space-borne experiments like GAIA and SIM, and 
could be within the reach of ground-based interferometers, such as 
PRIMA. Note that a periodic signal can be detected  with high confidence in a sufficiently long time series,  
even if the errors on the single measurements are larger than its amplitude \citep[see, e.g., ][]{HorneBaliunas86}. 
In the Sun, the average  lifetime of the sunspot groups dominating the flux modulation is $\sim 14$ days 
\citep[cf., e.g., ][]{Lanzaetal03,Lanzaetal07}, but they have a remarkable tendency to form in sequence
within activity complexes, the lifetime of which is $\sim 3-6$ months \citep[e.g., ][]{Gaizauskasetal83}. 

Therefore, if an activity complex on $\alpha$ Cen A  is stable for a time interval longer than the rotation period, 
it can produce a cyclic modulation of the position of the
photocentre with a semi-amplitude up to $\sim 8$ $\mu$as and a period of 29 days that can be erroneously 
attributed to the reflex motion
induced by a planet of $\stackrel{<}{\sim} 20$ Earth masses on an orbit of semi-major axis of $\sim 0.18$ AU, assuming that
the star has the same mass of the Sun. Such a spurious detection is difficult to disprove 
by correlating the photocentre motion with the
variation of the stellar optical flux, as measured by ground-based photometry, because the expected amplitude of 
variation is only $\sim 0.006$ mag (cf. Fig.~\ref{flux_l} and Eq.~\ref{flux_law}). 

On the average, the level of activity is expected to be higher than solar for the stars of spectral types later than the Sun
listed in Table~\ref{table1}, {
 because their convection zones are deeper and their rotation periods are comparable or shorter than the solar one 
\citep[cf., e.g., ][]{Weiss94}.
The expectation is confirmed  
 in the case of $\epsilon$ Eri, for which spot modelling of MOST photometry indicates a filling 
factor of at least 0.7\% \citep{Crolletal06}}. This implies a maximum excursion 
of $\sim 9$ $\mu$as for the photocentre position with the rotation
period of the star, i.e., $\sim 11.3$ days. Moreover, the observations indicate that the signal would maintain
its coherence for several rotations because the timescale of spot evolution is longer than 35 days. 
The observed amplitude of the optical flux variation of $\epsilon$ Eri is only $\sim 0.014$ 
mag, that is at the detection limit of ground-based photometry.  

It is interesting to note that $\epsilon$ Eri has a planetary companion on a wide orbit with a period of 
6.8 years \citep[e.g., ][]{Benedictetal06}. The astrometric orbit has a semi-major axis of $1.88 \pm 0.20$ mas
(milli arcseconds), that is much larger than the astrometric noise expected to be produced by stellar activity.
 The inclination of the planetary orbit is $i_{\rm p} = 30^{\circ}.1 \pm 3^{\circ}.8$.
Assuming that the stellar spin and the orbital angular momentum are aligned, we can adopt the same inclination for the stellar
rotation axis. This implies that all the active regions located at  latitudes higher than $30^{\circ}$ are circumpolar
and produce a continuous modulation of the photocentre position with the stellar rotation period. They can give rise to 
a spurious detection of further planetary companions around $\epsilon$~Eri. Specifically, assuming a stellar mass of
$0.83$ M$_{\odot}$, a photocentre oscillation with a period of $11.3$ days and
an amplitude of $9$ $\mu$as can be attributed to a planet of $\sim 0.3$ Jupiter masses on an orbit of
semi-major axis  $9.26 \times 10^{-2}$ AU.

Late-type stars with rotation periods of a few days have spot filling factors of the order of 
$f \sim 0.05-0.1$ { \citep[cf., e. g., ][]{Messinaetal03}}. They  
can produce astrometric signatures with amplitudes up to 0.1 stellar radii,
i.e., up to $\sim 30$ $\mu$as in the group of 66 nearby stars considered above. If we extend our sample to the
solar-type stars within 60 pc from the Sun, at least 13,000 objects brighter than $V =13$ are
expected to be observable by GAIA with an astrometric precision between 5 and 10 $\mu$as \citep{Sozzettietal01}.
The apparent radius of
the Sun at a distance of 60 pc is $78$ $\mu$as, implying that the most active stars in the sample (at least $2$\%$-3$\%) 
should display measurable astrometric effects at the level of $5-10$ $\mu$as. 
The  cyclic modulation of the signal with the rotation period
is much shorter than the typical timescales of starspot evolution, that range typically between $\sim 5$ and $\sim 16$ months
\citep[][]{MessinaGuinan03}. As a consequence, the astrometric modulation is quasi-stable, increasing the probability of
spurious planetary detections. Of course, the correlation of the astrometric signal with the photometric
variation can be used to discriminate between activity-induced and planetary-induced photocentre motions. Such stars would
have relative flux variations of the order of 0.1 mag,  that is about $30-100$ times larger than the photometric precision of GAIA.
However, the unknown value of the inclination of the stellar rotation axis leaves an ambiguity in those cases
in which no flux variation can be detected. 

\section{Discussion and conclusions}

We have pointed out the importance of taking into account the astrometric effects of surface brightness inhomogeneities
in future astrometric search for extrasolar planets, in particular of planets with an orbital period of 
a few days or a few tens of days around late-type stars. The simultaneous measurement of stellar wide-band fluxes
offers a powerful tool to discriminate activity-induced effects  from the reflex motion of the photocentre induced
by an orbiting body. However, the cases of nearby stars, such as $\alpha$ Cen or $\epsilon$ Eri, or 
of rotators observed almost pole-on 
require some additional methods to investigate the possible effects of magnetic activity. The fraction of stars
having an inclination lower than, say, $20^{\circ}$ is 6.0\%, which implies that at least a few tens of
ambiguous cases should be expected in the GAIA sample of stars within 60 pc. 

A further method of discrimination in those cases comes
from the fact that the activity-induced motion has the same period of the rotation of the star. If the inclination is not
too low, it can be measured from the modulation of the
chromospheric Ca II H \& K or Ca infrared triplet 
fluxes, even for stars with an activity level as low as the Sun \citep[e.g., ][]{Baliunasetal95,Andrettaetal05}, or 
can be roughly estimated from the spectroscopic $v \sin i$ and an assumed stellar radius.
Conversely, when the star is viewed almost pole-on, no rotational modulation is expected, but the level of 
activity can be estimated from those chromospheric indicators { as well as from the stellar X-ray flux.  

These considerations indicate the importance of a characterization of the level of activity and
the determination of the rotation period for those stars that have been selected as potentially planet-harbouring 
objects by means of the astrometric method.
This is particularly important when the measurement samples consist of widely separated data, as will be the 
case for GAIA that will observe a given star for 
about 80 times in five years, on the average.
Dedicated simulations, based on the observed positions of solar active regions along an entire activity
cycle and the time-sampling law of GAIA, are required to assess the impact of stellar activity on GAIA
long-term astrometry for solar-like stars. This will be the subject of a future work. Here we limit ourselves to
some general considerations based on solar analogy. 

Individual spot groups  will grow and decay on 
timescales of a few weeks, 
nevertheless the astrometric signal due to photospheric activity may show some phase coherence
for timescales comparable with the lifetimes of complexes of activity. In the case of sunspot
complexes, those timescales do not exceed $5-6$ solar rotations \citep{Gaizauskasetal83}. However, \citet{Schroeter84} 
reported the persistence of two chromospheric active regions separated by $\sim 180^{\circ}$ in longitude
for time intervals longer than $\sim 500$ days during periods of moderate or high solar activity. 
If a similar behaviour characterizes magnetic activity in other solar-like stars, this can give rise
to spurious planetary detections given that the astrometric signal produced by two active regions
separated by $\sim 180^{\circ}$ in longitude shows a sinusoidal oscillation with a period half that of
stellar rotation.  

Stars with a level of activity significantly higher than our Sun often show active longitudes
of spot activity lasting for decades 
\citep[see, e.g., ][]{Rodonoetal00,Rodonoetal01,Lanzaetal02,GarciaAlvarezetal03,Lanzaetal06} 
that can give rise to spurious astrometric detections of bodies orbiting around them.  
Modelling of the wide-band fluxes of highly active stars indicates that cool spots usually dominate their 
photometric variations. 
Moreover, extrapolations based on solar observations at different levels of activity along the 
11-yr cycle, support the conclusion that their active regions are
dominated by cool spots with bright faculae playing a secondary role 
\citep{Foukal98}. 
This implies that their activity-induced astrometric signal can be
 estimated from the first of equations (\ref{power_law}) (or the corresponding plot in Fig.~\ref{excur_l} for 
$f \geq 0.1$). 
}

\ack
AFL and CDM wish to dedicate this work to Marcello Rodon\`o, Professor of Astronomy in the University of
Catania and Director of the Astrophysical Observatory of Catania, who suddenly passed away during the 
preparation of the present paper. The authors wish to thank the anonymous Reviewers for their valuable comments. 
AFL wishes to thank Drs. M.~G.~Lattanzi and A.~Spagna for useful discussions. 

Active star research at INAF-Catania Astrophysical Observatory and the Department of Physics
and Astronomy of Catania University is funded by MIUR ({\it Ministero dell'Universit\`a e della Ricerca}), 
and by {\it Regione Siciliana}, whose financial support is gratefully
acknowledged.

This research has made use of the ADS-CDS databases, operated at the CDS, Strasbourg, France.




\end{document}